# Tunable magnetoresistance behavior in suspended graphitic multilayers through ion implantation

Carlos Diaz-Pinto, Xuemei Wang, Sungbae Lee, Viktor G. Hadjiev, Debtanu De, Wei-Kan Chu and Haibing Peng*

Department of Physics and the Texas Center for Superconductivity, University of Houston, Houston, Texas 77204, USA

* Corresponding author: haibingpeng@uh.edu

ABSTRACT

We report a tunable magnetoresistance (MR) behavior in suspended graphitic multilayers through point defect engineering by ion implantation. We find that ion implantation drastically changes the MR behavior: the linear positive MR in pure graphitic multilayers transforms into a negative MR after introducing significant short-range disorders (implanted boron or carbon atoms), consistent with recent non-Markovian transport theory. Our experiments suggest the important role of the non-Markovian process in the intriguing MR behavior for graphitic systems, and open a new window for understanding transport phenomena beyond the Drude-Boltzmann approach and tailoring the electronic properties of graphitic layers.

Despite intensive research on graphitic structures,[1-8] there remains a long standing problem regarding the origin of the intriguing linear positive magnetoresistance (MR) observed in bulk graphite [1,5] and recently in graphitic multilayers.[6,7] Such a strong linear MR is unexpected by the Drude-Boltzmann transport theory and thus drew immediate attention after its discovery.[5,8] Earlier attempts [8] were directed to its potential quantum origin at high magnetic fields where all carriers are condensed into the first Landau level (i.e. the "extreme quantum limit"), yet this deviates from the experimental observation that a linear MR appears also in $B$ fields far below this limit.[1] Recently, non-Markovian transport theory beyond the Drude-Boltzmann approach attracted considerable attention [9-20] in explaining the MR behavior of GaAs-based two-dimensional (2D) systems. Such a theory predicts a strong positive MR in a broad range of $B$ fields under the interplay of both short- and long-range disorders,[12,14] while a negative MR is expected with one type of disorder dominating (either long- or short-range).[15-17] Here we report the experimental evidence of a disorder-tunable MR behavior in suspended graphitic multilayers through point defect engineering by ion implantation. We find that ion implantation drastically changes the MR behavior as predicted by the non-Markovian transport theory: the linear positive MR in pure graphitic multilayers transforms into a negative MR after introducing significant short-range disorders (implanted boron or carbon atoms). This suggests that the origin of the unusual linear positive MR in pure graphitic structures is the non-Markovian transport under the interplay between long-range disorders (adsorbed charge impurities at the surface) and short-range disorders (defects or impurities inside the lattice). After ion implantation, short-range disorders dominate and lead to a distinct negative MR behavior. Our experiments open a new window not only for further understanding transport phenomena beyond the Drude-Boltzmann approach, but also for tailoring the electronic properties of graphitic layers for potential future applications.

In traditional Boltzmann theory, the disorder effect is characterized by completely uncorrelated scattering events. For transport with single type of carrier, this leads to the Drude-form longitudinal resistance independent of $B$.[21] (For transport with two types of carrier, the Boltzmann approach [1,21] predicts a positive MR showing a $B^2$ dependence at low $B$ and saturating at $\mu B \gg 1$ with $\mu$ the carrier mobility.) However, for 2D electrons moving under perpendicular $B$ fields, the cyclotron orbits result in a finite probability that an electron returns and collides again with the same scatterer at a later time. Such recollisions introduce memory effects between scattering events, and the non-Markovian approach has to be incorporated in describing 2D magneto-transport,[9-11] which leads to distinct MR behaviors dependent on the type of disorder. In the case of long-range disorders, the guiding center of a cyclotron orbit drifts along the constant-energy contours of the disorder potential under high magnetic fields (Fig. 1a), and electron localization occurs for all contours except for those rare percolating contours. This results in a negative MR behavior at high $B$ (i.e., a zero longitudinal resistance and a finite transverse resistance when localization occurs).[11-13] In the case of dominating short-range disorders (e.g. hard disk scatterers) (Fig. 1b), repeated scattering by the same scatterer also leads to localization and thus a negative MR behavior at high $B$.[9,10,13] On the contrary, in the presence of both long- and short-range disorders (Fig. 1c), cyclotron guiding centers drift mostly along the equipotential lines of the long-range disorder potential while the scattering by short-range disorder induces transitions between such equipotential contours, leading to electron delocalization and therefore a non-saturating positive MR even at high $B$ fields.[14] Our experiments are motivated by the prediction that the MR behavior strongly depends on the type of disorder, and we investigate the magneto-transport in suspended graphitic multilayers before and after disorder engineering through ion implantation.

We explore a distinct experimental procedure [22, 23] for the fabrication of devices containing suspended graphitic multilayers. Graphitic multilayers are suspended and electrically bridge the source and drain electrodes spaced ~ 500 nm apart (Fig. 1d). A typical device is shown in Fig. 1e. In experiments, we first characterize the magneto-transport properties of a suspended graphitic multilayer, followed by implantation of boron (or carbon). Finally, magneto-transport experiments are performed again on the same device after ion implantation to compare the MR behaviors. Before and after the ion implantation, the device is also characterized by Raman spectroscopy.

Figs. 2a and 2b present the differential conductance ($dI/dV$) as a function of back-gate voltage $V_g$ and drain-source bias $V_d$ at temperature $T = 4.2K$ for a graphitic multilayer with a thickness ~ 16 nm. The gate tunable $dI/dV$ reveals a conductance minimum related to the electronic band overlap in graphitic multilayers with an unintended *p*-type doping likely due to surface adsorbents, while the $V_d$ dependence of $dI/dV$ reflects the temperature dependence of $dI/dV$. [22,23]

Fig. 2c shows the differential resistance $R$, the inverse of measured $dI/dV$, as a function of $B$ field at different $V_d$. We note in passing that contact resistance does not obscure the study of intrinsic MR behavior for the graphitic multilayer since it merely adds an offset to the plotted $R$. Two notable features are observed. (1) For $V_d = 0$, the data demonstrate a nearly linear positive MR background with weak Shubnikov-de Haas (SdH) oscillations. But for higher $V_d$, the weak SdH oscillations are washed out (by a higher effective temperature) [23] and a nearly linear positive MR is revealed. (2) There is a clear negative MR at lower fields ($B < 1.2$ T) for $V_d = 0$ but its magnitude gradually diminishes as $V_d$ increases, which indicates that its physical origin is distinct from that for the nearly linear positive MR (persistent at high $V_d$ and non-saturating even at $B$ = 15 T). Such a negative MR at lower $B$ and its attenuation at higher $V_d$ (or higher effective temperature) is consistent with the weak localization picture, and the data can be well fitted with the theory for graphitic layers (inset of Fig. 2c).[24] We note that for some other devices [25] no appreciable weak localization was observed, but we present here a device showing significant negative MR at low $B$ in order to clearly distinguish the physical origins of the negative MR at low $B$ and the positive MR at high $B$.

Now we discuss the physical mechanism for the main feature: the nearly linear positive MR non-saturating even at $B$ = 15 T. A suspected linear positive MR background on top of SdH oscillations has been noted in graphitic structures for decades,[1,5-7] yet its physical origin is still in debate. An earlier attempt was directed to a quantum origin at high $B$-field where all carriers condense into the lowest Landau level,[8] but this deviates from the experimental observation of a linear MR over a $B$ field range even before the onset of SdH oscillations.[1,25] The persistence of a nearly linear positive MR at higher effective temperature (higher $V_d$ from Fig. 2c) also excludes a potential quantum origin suggested by Ref. 8 and points towards a quasi-classical mechanism.

Further, we argue that the origin of the MR behavior lies in the non-Markovian transport under the influence of both long- and short-range disorders (Fig. 1c).[14] For the graphitic multilayer in a geometry of Fig. 1d, the conduction channel is closer to the bottom surface due to parallel conductance in multilayers. Random charged adsorbents on the top surface lead to a long-range smooth disorder potential characterized by a correlation length $d$ close to the multilayer thickness (~ 16 nm for the device of Fig. 2). On the other hand, atomically sharp defects in graphitic multilayers and adsorbents on the bottom surface (if near the conduction channel) serve as short-range disorders. The interplay between long- and short-range disorders is predicted [14] to destroy the $B$-field-induced localization (Fig. 1c), and this can explain the non-saturating positive MR even at very large $B$ for graphitic multilayers. Considering short-range disorders as hard disks of radius $a$ (randomly placed with a concentration $n$), non-Markovian transport theory [14] predicts a nearly linear positive MR ($R \propto B^{0.77}$) at high $B$ fields in the hydrodynamic limit ($na$ = constant while $a \rightarrow 0$). Fig. 2c shows the best fittings by a power law behavior ($R \propto B^{\alpha}$) to the MR data above $B = 2.15$ T. For $V_d = 0$, the best fitting gives $\alpha \sim 0.81$, close to the theoretical value $\alpha \sim 0.77$. The obtained $\alpha$ values for higher $V_d$ are smaller. The $V_d$ dependence of $\alpha$ could be related to the dependence of the scattering cross section on the electron energy for a real impurity, which is not captured by the simple hard disk model [14] and requires future theoretical investigation.

As previously discussed, non-Markovian transport leads to a MR behavior determined by the type of disorder. To verify this, we performed disorder engineering through boron ion implantation [23] in the device of Fig. 2, and subsequently compared its MR behavior before and after the implantation. Raman spectra were also collected to provide structural information (Fig. 3a). The boron implantation (with a boron density per layer $n_S \sim 1.4 \times 10^{12}$ /cm$^2$) induced an increase of the intensity ratio between the $D$ and $G$ band ($I_D/I_G$), reflecting an increase of structural defects due to the implantation. [26,27] There is also an increase in both the G and D band frequency and intensity of the intra-valley disorder-induced D' band at ~1620 cm$^{-1}$. A brief discussion of these effects is given in Ref. 28. We estimate, based on Ref. 29, the in-plane crystal size $L_a = 16.6$ ($I_G/I_D$) ~ 10 nm after the boron implantation, while $L_a$ is ~ 41 nm before the implantation. Therefore, the Raman spectra confirm that a significant amount of disorders are introduced by the implantation of boron, which was classified [27] as a type of short-range disorder.

We now address the electron transport for the same device after the introduction of short-range disorders (boron atoms) in the graphitic lattice. First, the boron implantation introduces *p*-type doping, as shown by the gate tuning behavior (Fig. 3b). The p-type behavior is clear in the data for $T = 40$ K, while it is masked by conductance oscillations at lower $T$ due to the interference of confined electron waves. Second, significant differences appear in the data of $dI/dV$ vs. $V_d$ (Fig. 3c), which reflect a change of the temperature dependence of $dI/dV$ after the boron doping. [22, 23] Before the implantation, the $dI/dV$ displays a shallow dip near $V_d = 0$, followed by a significant rise of $dI/dV$ even up to $V_d = 200$ mV. After the boron implantation, the $dI/dV$ value at $V_d = 0$ is higher due to the carrier doping effect and there is also a $dI/dV$ dip near $V_d = 0$. While at higher $V_d$, the slope for the $dI/dV$ increase is much smaller compared with that before the implantation. The observed $V_d$ dependence of $dI/dV$ (Fig. 3c) can be attributed to a temperature effect due to electron or lattice heating at finite $V_d$, as indicated by a similar shape in the temperature dependence of $dI/dV$ at $V_d = 0$ measured in experiments. [23]

The most striking feature is the drastic change of MR behaviors: the boron implantation results in a negative MR (Fig. 4), in stark contrast to the positive MR observed before the implantation (Fig. 2c). After the implantation (Fig. 4), for $B$ lower than 6 T the negative MR follows a parabolic $B$ dependence, while for higher $B$ it shows a linear dependence. We stress that such a switching from a positive MR to a negative MR behavior is reproducible in experiments, and we have similar results for five samples after ion implantation with a density of short-range scatters $n_S \sim 10^{12}$ /cm$^2$ per layer. (We also observed the switching of MR behavior with carbon implantation [23], which suggests that the MR behavior is determined by the disorder effect instead of the carrier doping effect since the carbon ion does not introduce hole doping as boron does.) In a device with lower implanted boron density $n_S \sim 10^{10}$ /cm$^2$ per layer, the slope of the positive MR is decreased but a crossover to negative MR is not reached. [23] For devices with higher boron density $n_S > 10^{14}$ /cm$^2$ per layer, we observe nearly insulating behavior after implantation, suggesting appearance of strong localization.

According to non-Markovian transport theory,[13] a parabolic negative MR appears when strong short-range disorders dominate over long-range smooth disorders, resulting from the memory effect induced by the return of cyclotron orbits to the starting points after one or more revolutions. Our experimental observation of the switching from a positive MR to a negative MR confirms the non-Markovian picture since significant short-range disorders (boron or carbon atoms) are introduced by the implantation. Fig.4 shows a fitting of the data with a parabolic MR [13]: $R(B) = R_0[1-(\omega_c/\omega_0)^2]$, where $R_0$ is the resistance at $B = 0$, $\omega_c = eB/m^*$ is the cyclotron frequency with $m^*$ the effective mass, and $\omega_0 = (2\pi n_S)^{1/2} v_F (2\gamma l_S/l_L)^{1/4}$ is a constant frequency with $n_S$ the density of short-range scatterers, $v_F$ the Fermi velocity, $l_S$ ($l_L$) the transport mean free path due to short- (long-) range disorders, and $\gamma \sim 1$ a constant parameter. Taking $m^* = 0.039 m_e$ for holes in graphite, we obtain $\hbar\omega_0 = 156$ meV from the parabolic fitting. For $B = 6$ T, $\hbar\omega_c = 17.8$ meV is much less than $\hbar\omega_0$, which justifies the experimentally observed parabolic negative MR up to 6 T according to Ref. 13. Also, from the ion dosage in experiments we estimate the density of short-range scatters (boron ions) $n_S \sim 1.4 \times 10^{12}$ /cm$^2$ per layer near the conduction channel, and thus from the fitting value of $\omega_0$ we obtain a ratio $l_L/l_S \sim 6$ by taking the graphene Fermi velocity $v_F = 1 \times 10^6$ m/s. This again confirms that short-range disorders dominate over long-range disorders after the boron implantation, as expected. (Assuming boron ions as hard wall scatterers with a radius $a \sim 0.5$ nm as reported previously,[27] we can estimate $l_S \sim 1/2n_S a =$ 70 nm and $l_L \sim 420$ nm.) For *B* fields higher than 6 T, we find a linear negative MR (Fig. 4), which is not quantitatively understood at the moment.

    In summary, comparing the magneto-transport in graphitic multilayer before and after ion implantation, we find drastically different MR behaviors determined by the disorder effect, as predicted by non-Markovian transport theory. Our work therefore suggests the important role of the non-Markovian process in the intriguing MR behavior for graphitic systems.

# REFERENCES

1. I. L. Spain, *Carbon* **17,** 209 (1979).
2. K.S. Novoselov, *Nature* **438,** 197 (2005).
3. Y. Zhang, Y. W. Tan, H. L. Stormer, P. Kim, *Nature* **438,** 201 (2005).
4. A.H. Castro Neto, F. Guinea, N.M.R. Peres, K.S. Novoselov, A.K. Geim, *Rev. Mod. Phys.* **81**, 109 (2009).
5. D.E. Soule, *Phys. Rev.* **112,** 698 (1958).
6. Y. Zhang, J.P. Small, M.E.S. Amori, P. Kim, *Phys. Rev. Lett.* **94,** 176803 (2005).
7. K.S. Novoselov et al., *Science* **306**, 666, (2004).
8. J.W. McClure, W.J. Spry, *Phys. Rev.* **165,** 809 (1968).
9. A.V. Bobylev, F.A. Maao, A. Hansen, E.H. Hauge, *Phys. Rev. Lett.* **75,** 197 (1995).
10. E.M. Baskin, M.V. Entin, *Physica B* **249-251,** 805 (1998).
11. M.M. Fogler, A.Y. Dobin, V.I. Perel, B.I. Shklovskii, *Phys. Rev. B* **56,** 6823 (1997).
12. A.D. Mirlin, J. Wilke, F. Evers, D.G. Polyakov, P. Wolfle, *Phys. Rev. Lett.* **83,** 2801 (1999).
13. A.D. Mirlin, D.G. Polyakov, F. Evers, P. Wolfle, *Phys. Rev. Lett.* **87,** 126805 (2001).
14. D.G. Polyakov, F. Evers, A.D. Mirlin, P. Wolfle, *Phys. Rev. B* **64,** 205306 (2001).
15. A. Dmitriev, M. Dyakonov, R. Jullien, *Phys. Rev. B* **64,** 233321 (2001).
16. A. Dmitriev, M. Dyakonov, R. Jullien, *Phys. Rev. Lett.* **89,** 266804 (2002).
17. V.V. Cheianov, A.P. Dmitriev, V.Y. Kachorovskii, *Phys. Rev. B* **68,** 201304(R) (2003).
18. Z.D. Kvon, V. Renard, G.M. Gusev, J.C. Portal, *Physica E* **22**, 332 (2004).
19. N.M. Sotomayor *et al.*, *Phys. Rev. B* **70,** 235326 (2004).
20. H.I. Cho *et al.*, *Phys. Rev. B* **71,** 245323 (2005).
21. J.M. Ziman, "Principles of the theory of solids" (2$^{nd}$ Ed.), Cambridge University Press, 1972.
22. S. Lee, N. Wijesinghe, C. Diaz-Pinto, H.B. Peng, *Phys. Rev. B* **82,** 045411 (2010).
23. See Supplementary Materials [Document # ] for experimental details, the temperature dependence of *dI/dV*, SdH oscillations, and the MR behavior for a device with carbon implantation and a device with lower density boron implantation.
24. E. McCann *et al.*, *Phys. Rev. Lett.* **97,** 146805 (2006).
25. C. Diaz-Pinto, S. Lee, V.G. Hadjiev, H.B. Peng, arXiv:1004.2987.
26. T. Hagio, M. Nakamizo, K. Kobayashi, *Carbon* **27,** 259 (1989).
27. M. Endo, T. Hayashi, S.H. Hong, T. Enoki, M.S. Dresselhaus, *J. Appl. Phys.* **90,** 5670 (2001).
28. We suggest that at least two mechanisms contribute to the *G* band frequency change with ion implantation. Firstly, boron doping can produce a sizable blue shift through the non-adiabatic Kohn effect on the Γ phonons.[30] Calculations according to Ref. 30 show that a shift of 5.5 cm$^{-1}$ corresponds to a doping of 3.5 x 10$^{12}$ holes/cm$^2$, which is of the same order as what we expect from a substitutional boron density $n_S$ ~ 1.4 x 10$^{12}$ /cm$^2$ per layer in experiments. (Kohn effect on the K phonons can also induce the observed D band shift of 3.6 cm$^{-1}$.) On the other hand, upon doping the G mode linewidth in graphene is expected to sharpen.[30] The observed *G* band broadening (Fig. 3a inset) is then related to inhomogeneous broadening due to variation of the G mode frequency around boron atoms and carbon clusters. The latter contribution to the G mode frequency corroborates also with the intensity increase of the defect-induced, intra-valley, D' band at ~1620 cm$^-$

   [1].[31]
29. L.G. Cançado *et al.*, *Appl. Phys. Lett.* **88,** 163106 (2006).
30. M. Lazzeri, F. Mauri, *Phys. Rev. Lett.* **97**, 266407 (2006).
31. M. A. Pimenta, G. Dresselhaus, M.S. Dresselhaus, L. G. Cancado, A. Jorio, R. Saito, *Phys. Chem. Chem. Phys*. **9**, 1276 (2007).

FIGURE CAPTIONS

Figure 1. Diagrams of cyclotron motion in 2D system in perpendicular $B$ fields and device schemes for this work. (a) Guiding center drift of cyclotron orbits along the constant-energy contours of a long-range smooth disorder potential, which leads to localization at high $B$ fields. (b) Localization of electrons by repeated scattering with the same hard-disk (short range) scatterer. (c) Transition of cyclotron motion between equipotential contours in the presence of both long-range disorders and short-range (hard disk) scatterers, leading to electron delocalization. (d) Cross section (upper panel) and top view (lower panel) of the device geometry illustrating a graphitic multilayer suspended ~110 nm above the $SiO_2$/Si substrate. (e) Scanning electron microscope image of a typical device with a suspended graphitic multilayer bridging two electrodes spaced ~ 500 nm apart. Scale bar: 1 μm.

Figure 2. Electron transport behavior of a multilayer graphitic device before boron implantation. (a) Differential conductance ($dI/dV$) as a function of gate voltage $V_g$ with drain-source bias $V_d = 0$ for a graphitic multilayer at temperature $T = 4.2$ K and magnetic field $B = 0$. (b) $dI/dV$ vs. $V_d$ with $V_g = 0$ at $T = 4.2$ K and $B = 0$. (c) Differential resistance $R$ (symbols), the inverse of measured $dI/dV$, as a function of $B$ field at various source-drain bias $V_d$ at $T = 4.2$ K. Lines are the best fittings to the experimental data by a power law behavior ( $R \propto B^\alpha$ ) for $B > 2.15$ T. The fitting parameter is $\alpha$ = 0.81, 0.65, 0.47, 0.52, 0.60, 0.64, for the data with $V_d$ = 0, -8, -16, -24, -32, -40 mV, respectively (from top to bottom). The vertical line is a guide to the eye at $B = 2.15$ T. Inset: Low field negative magnetoresistance data for $V_d = 0$ (symbols) and the best fitting (line) according to the weak localization theory for graphene in Ref. 24 with fitting parameters: the dephasing length $L_\varphi = 68$ nm, the intervalley scattering length $L_i = 50$ nm, and a scattering

length $L_* = 1$ nm characterizing the trigonal warping effect and chirality-breaking elastic intravalley scattering.

Figure 3. Raman scattering spectra and electron transport behavior for the device of Fig. 2 after a boron implantation (with a boron density per layer $n_S \sim 1.4 \times 10^{12}$ /cm$^2$). (a) Raman spectra after (red) and before (black) the implantation. Inset: a zoom-in view of the $G$ band, showing a shift of $\Delta\omega_G = 5.5$ cm$^{-1}$. (b) $dI/dV$ vs. $V_g$ with $V_d = 0$ at different temperatures after the implantation. (c) Comparison of $dI/dV$ vs. $V_d$ with $V_g = 0$ after (red) and before (black) the implantation.

Figure 4 Differential resistance $R$ (symbols) as a function of $B$ field with $V_d = 0$ at $T = 4.2$ K for the device in Fig. 3 after the boron implantation. The solid line and the dashed line show fittings with a quadratic and a linear $B$ dependence for the MR, respectively.

FIGURES

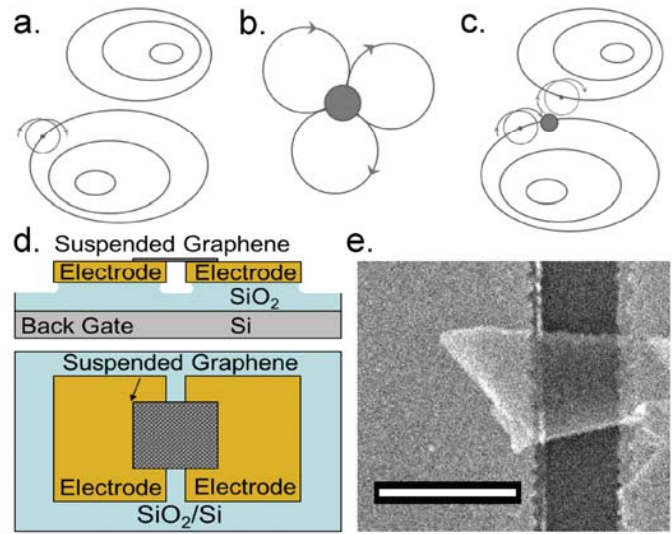

Figure 1

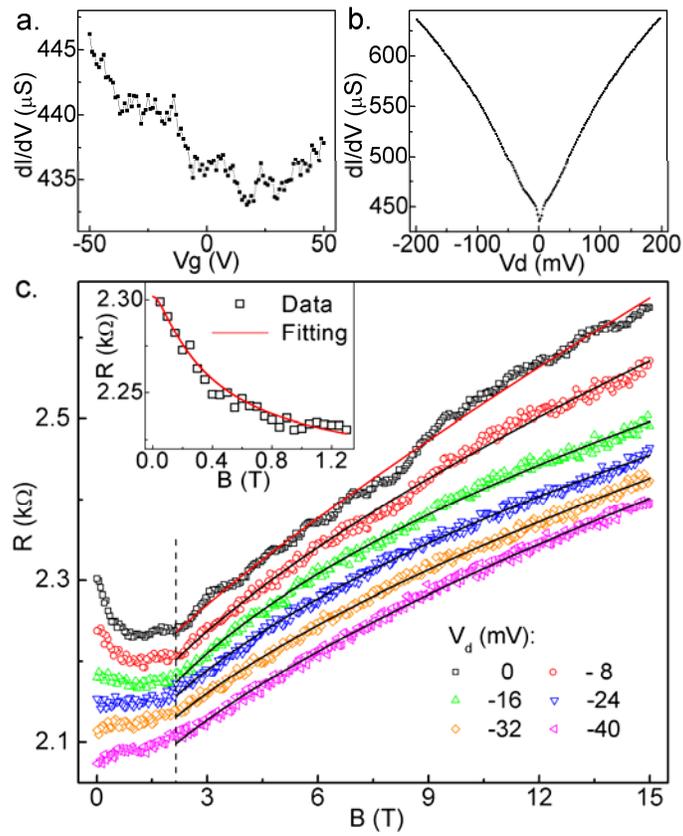

Figure 2

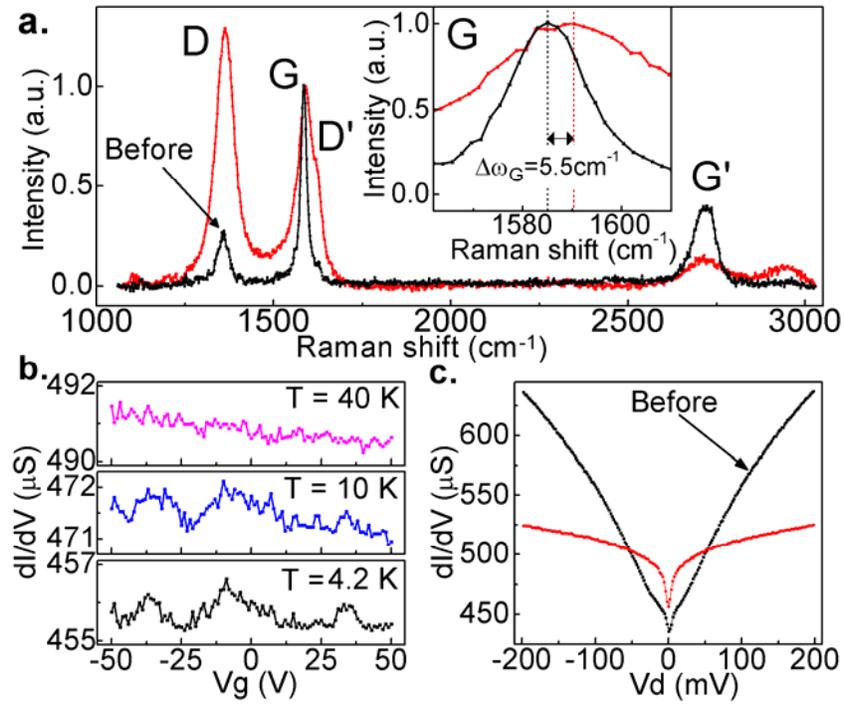

Figure 3

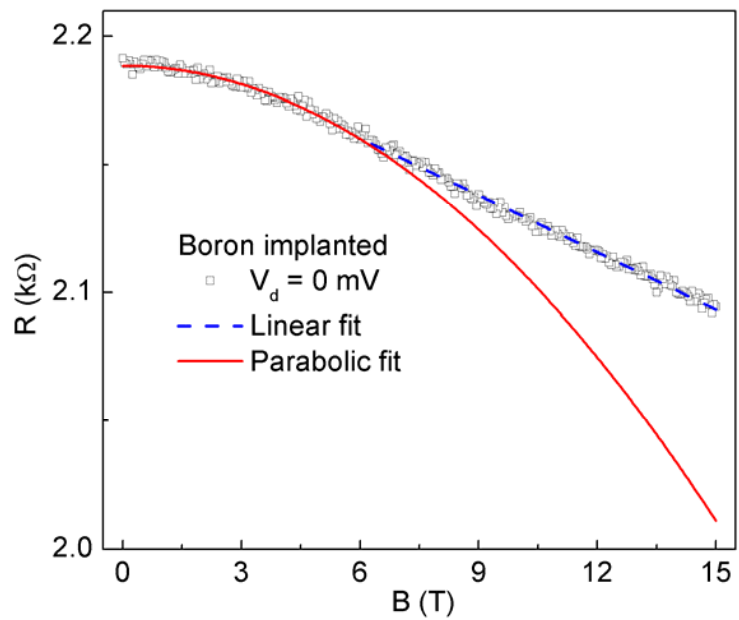

Figure 4

Supplementary Materials for:

**Tunable magnetoresistance behavior in suspended graphitic multilayers through ion implantation**

Carlos Diaz-Pinto, Xuemei Wang, Sungbae Lee, Victor G. Hadjiev, Debtanu De, Wei-Kan Chu and Haibing Peng*

Department of Physics and the Texas Center for Superconductivity, University of Houston, Houston, Texas 77204, USA

* Corresponding author: haibingpeng@uh.edu

## I. Experimental Details

Devices with suspended graphitic multilayers are constructed by the approach developed in Ref. 22 of the main text. The key point for device fabrication is to prepare liquid solutions containing a sufficient amount of thin multilayer graphitic flakes by mechanically grinding highly-oriented pyrolytic graphite. Source-drain electrode pairs separated by a 500 nm gap are first patterned by e-beam lithography on top of a degenerately doped Si wafer coated with a 200 nm layer of thermally grown $SiO_2$. Buffered hydrofluoric acid is then used to etch away 50 nm of $SiO_2$ so that the top surface of the electrodes (45 nm Pd/ 15 nm Cr) is ~110 nm higher above the basal oxide level. After that, multilayer graphitic flakes contained in liquid solutions are spin-coated onto the substrate with patterned electrodes. By chance, a graphitic flake is suspended between the elevated source and drain electrodes to form the three-terminal device (Fig. 1d of the main text). The relatively narrow gap between elevated electrodes enables us to obtain such devices with suspended graphitic flakes electrically connecting the source and drain. Subsequently, the devices are annealed at a temperature of 200 ºC in forming gas (95 % of $Ar_2$ and 5% of $H_2$). Electrical characterization is then carried out in a three terminal configuration by using the degenerately doped silicon as a back gate. (In experiments, the gate voltage $V_g$ was limited to ± 50 V to avoid the mechanical collapse of suspended graphitic flakes.) The $dI/dV$ is measured by standard lock-in technique with a small excitation voltage (300 µV or less) at a frequency 503 Hz, superimposed to a DC bias $V_d$. Magnetic fields are applied perpendicular to the graphitic flakes with a superconducting magnet inside a cryostat.

Raman scattering was performed on graphitic flakes before and after ion implantation using a laser with a 514 nm wavelength (2.41 eV). Ion implantation was carried out with 7 keV boron ions from a source of negative ions by cesium sputtering (SNICS) of 1.7 MV Tandem Accelerator. We used commercial natural boron powder with 19.6% $^{10}$B and 80.4% $^{11}$B as the target source. $^{11}$B was selected by a 30° analyzing magnet. The ion current was measured with a Faraday cup located at the low energy end of the accelerator. For the sample studied in Figs. 2-4 of the main text, a dosage of $1.2\times10^{14}$ ions/cm$^2$ was used. According to the SRIM simulation, we estimate a boron density per layer $n_S \sim 1.4 \times 10^{12}$ /cm$^2$ for the bottom layer in this sample (~ 16 nm thick as measured by atomic force microscope).

## II. Temperature Dependence of the Differential Conductance

Fig. 1S Differential conductance $dI/dV$ as a function of drain-source bias $V_d$ with gate voltage $V_g = 0$ at different temperatures (a) before and (b) after the boron implantation for the device of Fig. 2 in the main text; $dI/dV$ as a function of temperature with $V_g = 0$ at different $V_d$ (c) before and (d) after the boron implantation.

The $V_d$ dependence of $dI/dV$ at temperature $T = 4.2$ K displays a shape similar to its temperature dependence at $V_d = 0$, suggesting that the $dI/dV$ vs. $V_d$ data reflect the effective temperature increase due to electron and lattice heating. (For lower $V_d$, electron heating is more likely due to a weak electron-acoustic phonon scattering, while at higher $V_d$, lattice heating is possible due to the increased scattering of electrons from higher energy phonons.) For $T < 20$ K, the curves of $dI/dV$ vs. $V_d$ are nearly overlapping except for the dip near $V_d = 0$ which can be explained by the hot electron effect.

It is notable from (c) and (d) that for lower $V_d$, the temperature dependence of $dI/dV$ shows semiconducting behavior, i.e., higher conductance at higher $T$ due to the increase of thermally excited carriers. However, for higher $V_d$, the temperature dependence of $dI/dV$ changes to a metallic behavior (i.e. $dI/dV$ decreases as $T$ increases), suggesting that the temperature dependence is dominated by the enhanced scattering of hot electrons at higher temperatures (e.g., by phonons for which the population increases at higher $T$).

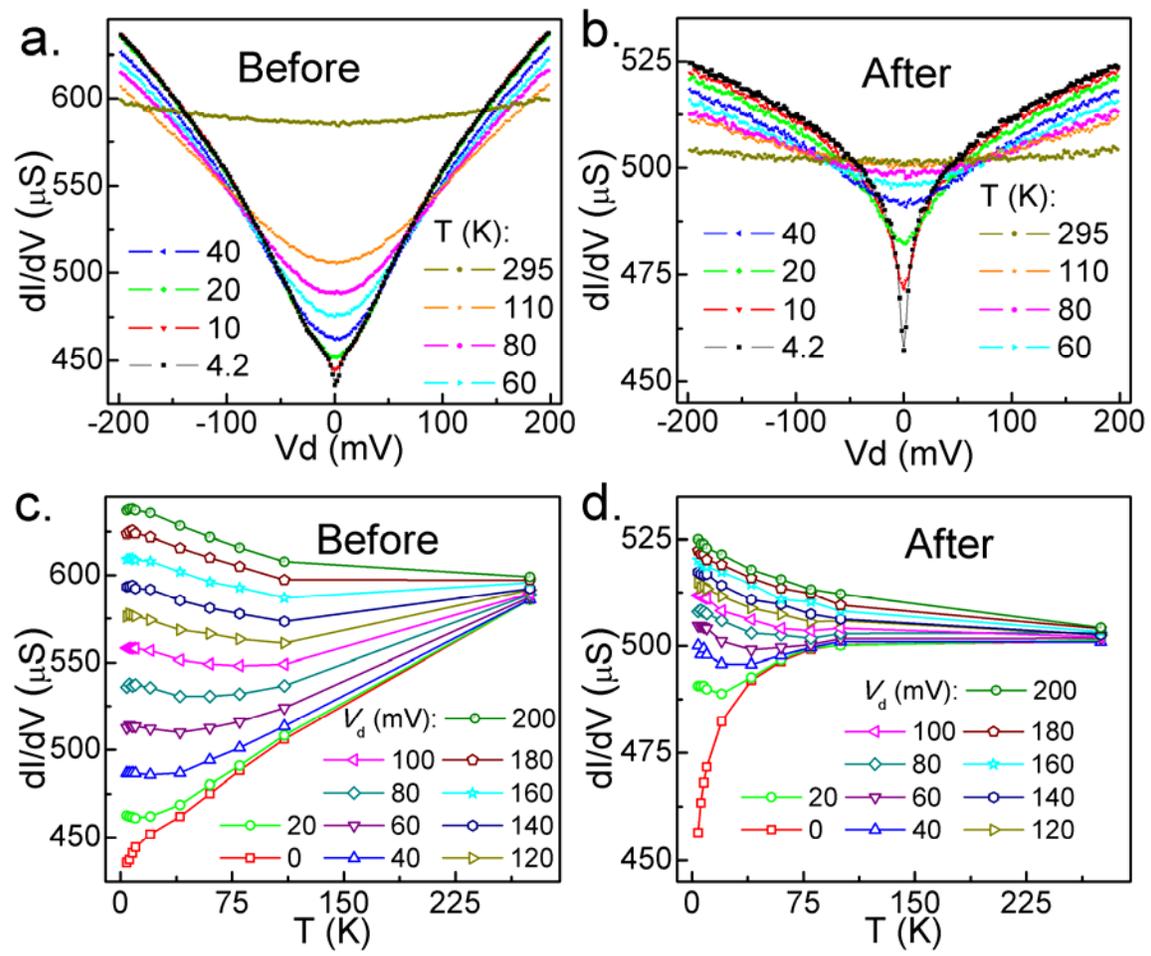

Fig. 1S

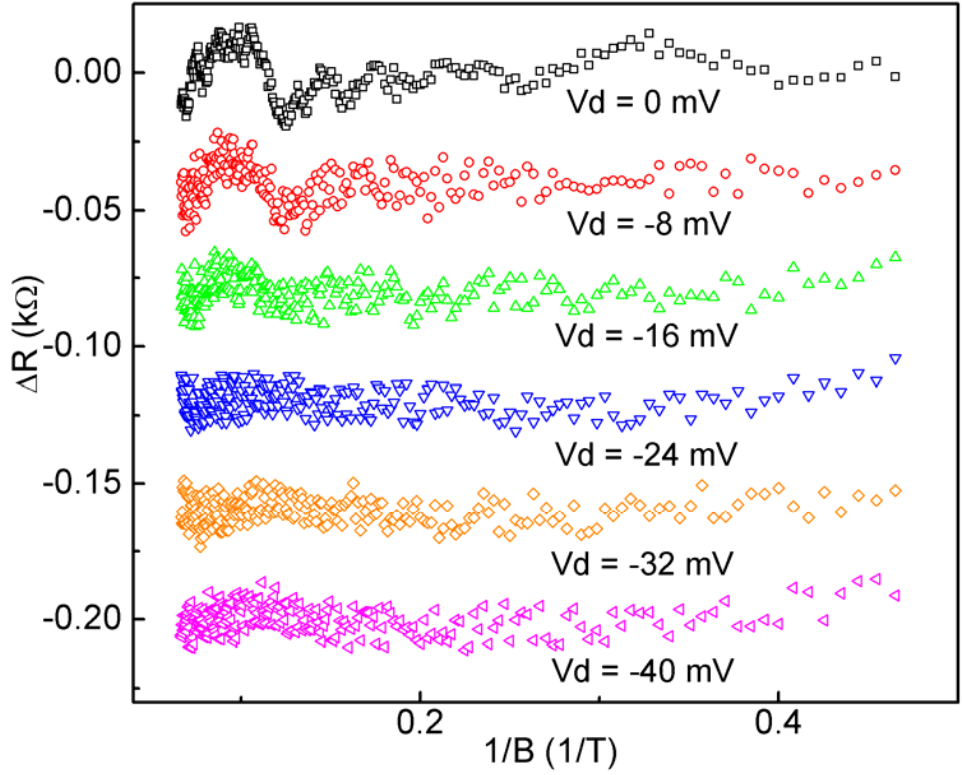

Fig. 2S

## III. Shubnikov-de Haas (SdH) oscillations

Fig. 2S SdH oscillations for the device of Fig. 2 in the main text at different drain source bias $V_d$ obtained by subtracting the original data with the fitted positive magnetoresistance (MR) background (Fig. 2c of the main text). For clarity, the curves are stacked with an offset of -0.04 kΩ from top to bottom.

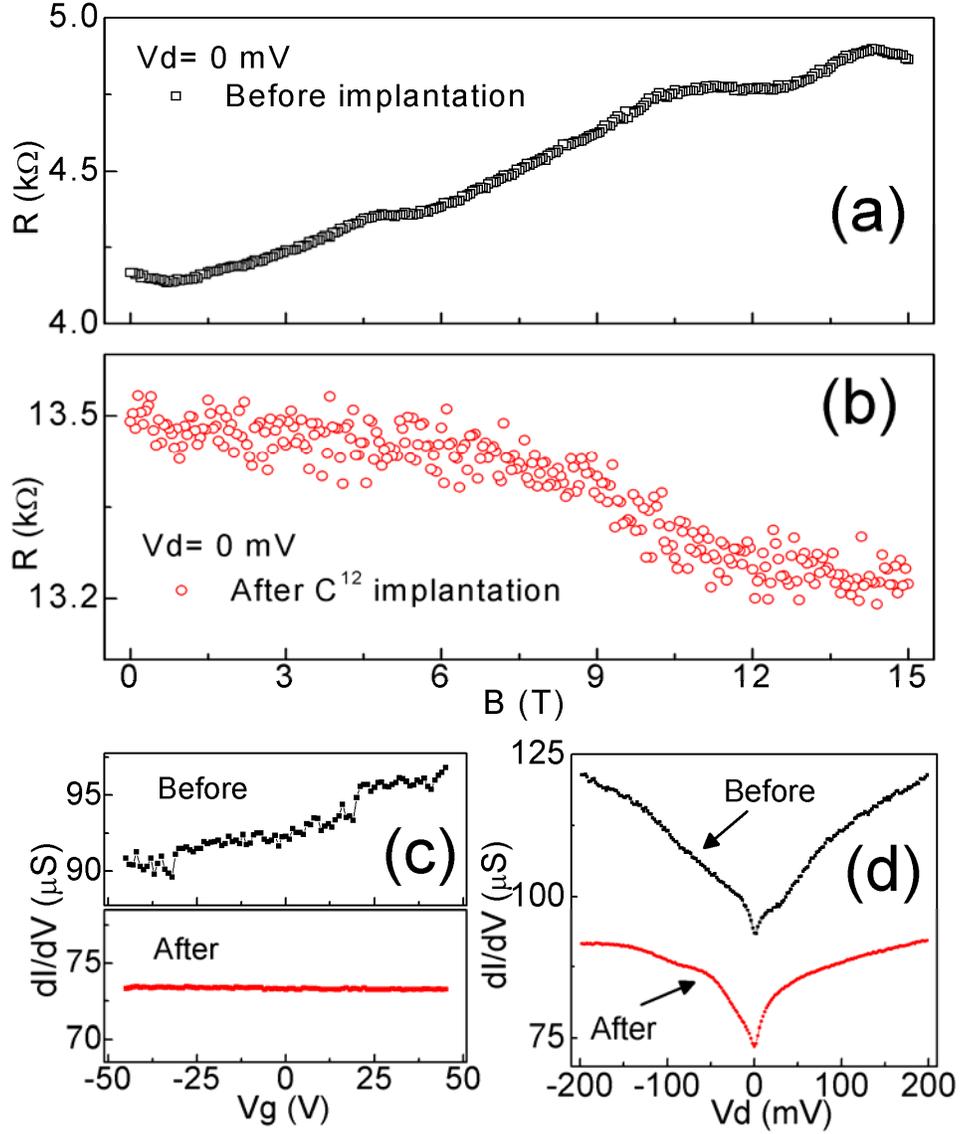

Fig. 3S

## IV. Magnetoresistance behavior induced by carbon implantation

Fig. 3S Magnetoresistance behavior for a suspended graphitic multilayer device (a) before and (b) after the implantation of $C^{12}$ ions (8 keV) with a total dosage of $1.2\times10^{14}$ ions/$cm^2$ in experiments (the implanted $C^{12}$ density per layer cannot be estimated since unfortunately we were not able to determine the thickness for this sample). (c) *dI/dV vs. $V_g$* and (d) *dI/dV vs. $V_d$* before and after the implantation for the device.

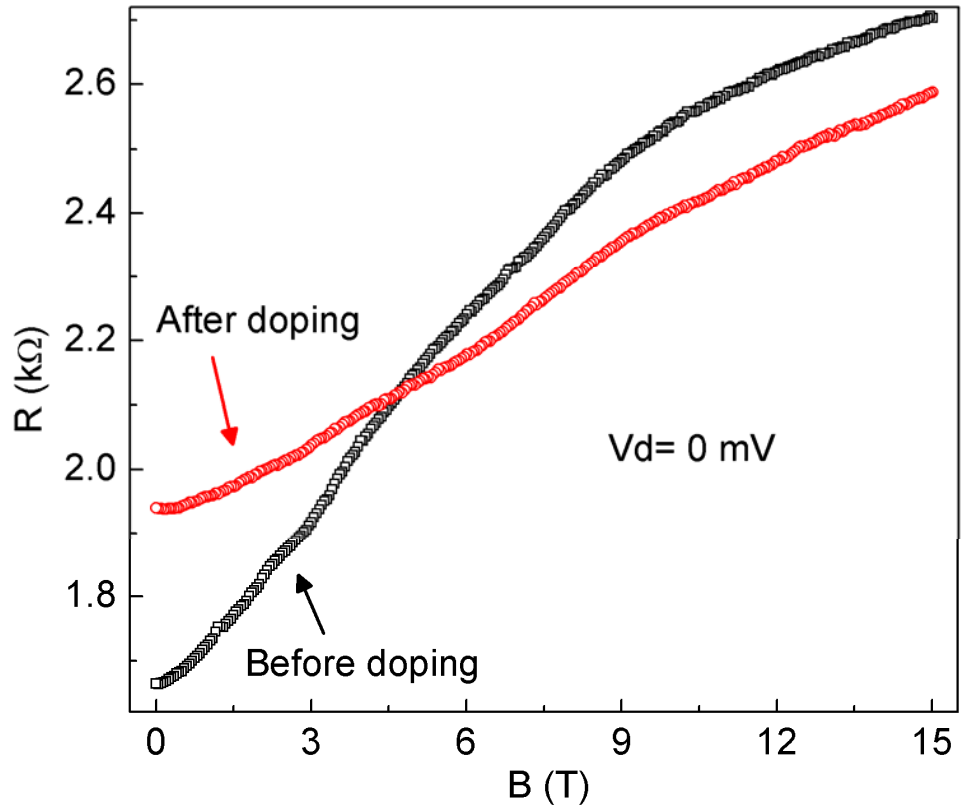

Fig. 4S

## V. Magnetoresistance behavior induced by lower density boron implantation

Fig. 4S  Magnetoresistance behavior for a suspended graphitic multilayer device (~ 49 nm thick) before (black) and after (red) the implantation of boron ions with a lower density ~ $10^{10}$ /cm$^2$ per layer for the bottom layer (according to the SRIM simulation with a dosage of $1.2\times10^{14}$ ions/cm$^2$ for implantation with 7 keV boron ions in experiments).